\documentclass{elsart}
\usepackage{epsf}

\begin{document}
\begin{frontmatter}

\title{Components of multifractality in high-frequency stock returns}

\author{J.~Kwapie\'n$^1$, P.~O\'swi\c ecimka$^1$ and 
S.~Dro\.zd\.z$^{1,2}$}

\address{$^1$Institute of Nuclear Physics, Polish Academy of Sciences,
PL--31-342 Krak\'ow, Poland \\ $^2$Institute of Physics, University of
Rzesz\'ow, PL--35-310 Rzesz\'ow, Poland}

\begin{abstract}

We analyzed multifractal properties of 5-minute stock returns from a
period of over two years for 100 highly capitalized American companies.
The two sources: fat-tailed probability distributions and nonlinear
temporal correlations, vitally contribute to the observed multifractal
dynamics of the returns. For majority of the companies the temporal
correlations constitute a much more significant related factor, however.

\end{abstract}

\end{frontmatter}

The first, simple stock market model by Bachelier~\cite{bachelier}, based
on Gaussian random walk, although it used to be applied for a long time,
did not successfully pass practical tests as soon as large data samples
were available and it was eventually rejected. The non-Gaussian
distributions of price fluctuations, their persistent nonlinear temporal
correlations, the intermittent behaviour present on all time scales and
even the explicitly shown scaling invariance, together with some other
properties of the stock market evolution, required introduction of more
advanced and more appropriate
models~\cite{mandelbrot68,mantegna95,plerou99,gopikrishnan99,plerou00,drozdz03}.  
What was especially interesting was the apparent similarity of the stock
market dynamics to fluid turbulence~\cite{ghasghaie96}, which led to the
development of models based on the multiplicative
cascades~\cite{mandelbrot97,calvet97,lux03a,lux03b,eisler04}. This sort of
processes generate signals which are inherently multifractal with a
continuous spectrum of scaling indices
$f(\alpha)$~\cite{halsey86,barabasi91}. Consistently, the real data from
different financial markets (stock, forex and commodity ones) show clear
multifractal
properties~\cite{pasquini99,ivanova99,bershadskii03,matteo04,fisher97,%
vandewalle98,bershadskii99,matia03,oswiecimka04}.

It has already been pointed out in literature that the two fundamental
factors leading to multifractal behaviour of signals are the nonlinear
time correlations between present and past events and the fat-tailed
probability distributions of fluctuations. Their role was analyzed both in
models~\cite{nakao00,kalisky04} and in the real data from
markets~\cite{matia03}. The latter work brought especially interesting
results from a perspective of the present study: based on the stock and
the commodity markets, authors of ref.~\cite{matia03} showed that, for the
stocks, the main contribution to multifractality comes from a broad
distribution of returns while a long memory present in this kind of data
contributes only marginally. In opposite, for commodities the reported
contribution of correlations is larger than for stocks. It should be noted
that the nature of correlations leading to the multifractal dynamics of
the returns is strongly nonlinear and, curiously, cannot be simply related
to some well-known correlation type like a slowly decreasing volatility
autocorrelation with an imposed daily pattern. For example, one even has
to consider the nonlinear dependencies in the volatility itself in order
to reveal how the temporal correlations contribute to multifractality in
the stock market data.

A widely used method of quantifying multifractal properties of time series 
is the (multifractal) detrended fluctuation analysis 
(MF-DFA)~\cite{peng94}. For a time series of $N_s$ logarithmic returns 
$g_s(i)=\ln p_s(t_i+\Delta t)-\ln p_s(t_i)$ of a stock $s$ (where 
$p_s(t_i)$ stands for price at discrete time $t_i$), one calculates the 
signal profile
\begin{equation}
Y(i) = \sum_{k=1}^i{(g_s(k)-<g_s>)}, \ i = 1,...,N_s
\end{equation}
where $<...>$ denotes averaging over the whole time series. Then $Y(i)$ is 
divided into $M_s$ disjoint segments of length $n$ ($n < N_s$) starting 
from both the beginning and the end of the time series so that finally one 
has $2 M_s$ such segments. For each segment $\nu$, the local trend has to 
be calculated by fitting a $l$-th order polynomial $P_{\nu}^{(l)}$ to the 
data. In the next step one calculates the variance for all $\nu$'s and 
$n$'s
\begin{equation}
F^2(\nu,n) = \frac{1}{n} \sum_{j=1}^n \{Y[(\nu-1) n+j] -
P_{\nu}^{(l)}(j)\}^2
\label{std.dev}
\end{equation}
and average it over $\nu$'s, obtaining the $q$th order fluctuation 
function
\begin{equation}
F_q(n) = \bigg\{ \frac{1}{2 M_s} \sum_{\nu=1}^{2 M_s} [F^2(\nu,n)]^{q/2}
\bigg\}^{1/q}, \ \ q \in \mathbf{R}
\end{equation}
for all choices of $n$. For a signal with fractal properties, the 
functional dependence $F_q(n)$ reveals power-law scaling
\begin{equation}
F_q(n) \sim n^{h(q)}
\label{scaling}
\end{equation}
for sufficiently large $n$. The final product of the MF-DFA procedure is 
a family of scaling exponents $h(q)$ (generalized Hurst exponents) which 
for an actual multifractal signal form a decreasing function 
of $q$ (for monofractals $h(q)=const$). From the spectrum of generalized 
Hurst exponents one can calculate the singularity spectrum $f(\alpha)$ by 
using the following relations:
\begin{equation}
\alpha=h(q)+q h'(q) \hspace{1.0cm} f(\alpha)=q [\alpha-h(q)] + 1.
\label{singularity}
\end{equation}

\begin{figure}
\hspace{0.0cm}
\epsfxsize 14cm
\epsffile{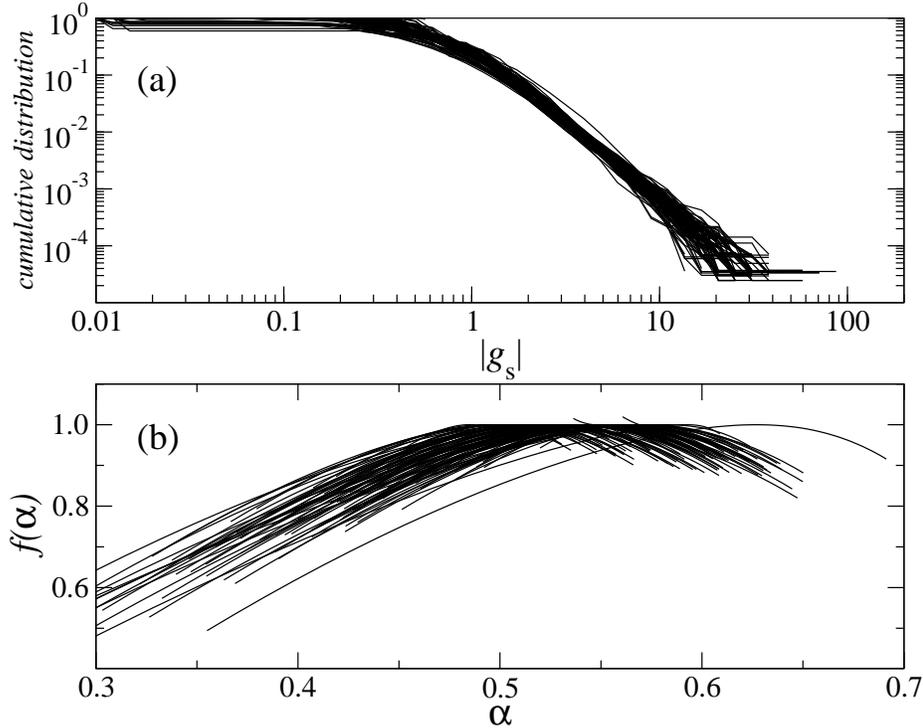}
\caption{(a) The cumulative distribution functions of normalized stock
returns for the 100 companies under study and (b) their singularity
spectra $f(\alpha)$.}
\label{individual}
\end{figure}

We analyzed time series of 5-minute stock returns for the 100 largest
American companies\footnote{The companies were: AA ABT AHP AIG ALD AMGN
AOL ARC AT AUD AXP BA BAC BEL BK BLS BMY BUD C CA CBS CCL CCU CHV CL CMB
CMCS COX CPQ CSCO DD DELL DH DIS DOW EDS EK EMC EMR ENE F FBF FNM FRE FTU
G GCI GE GLW GM GTE GTW HD HWP IBM INTC JNJ JPM KMB KO LLY LOW LU MCD MDT
MER MMC MMM MO MOT MRK MSFT MTC MWD ONE ORCL PEP PFE PG PNU QCOM QWST SBC
SCH SGP SLB SUNW T TWX TX TXN UMG USW UTX VIAB WCOM WFC WMT XON YHOO.}
according to their capitalization at the end of 1999. All the stocks were
traded on NYSE or NASDAQ. The time interval covered by our data was from
Dec 1, 1997 to Dec 31, 1999. We chose $\Delta t=5$ min because our time
series of about 40,000 points were long enough to obtain statistically
significant results and also length of the longest intervals of a constant
stock price (i.e. zero returns), which affect the results of the MF-DFA
procedure, was reasonably small and didn't exceed 30 points. In order to
eliminate a possible bias of the results due to the largest fluctuations,
we restricted the range of $q$ in MF-DFA to the interval (-3,3); this
restriction is also desired because of the inverse cubic power law which
practically rules out the existence of moments higher than for $q=3$.  
The companies which we analyzed represented various market sectors,
various trading frequencies (0.01-1 transactions/s), and capitalization
spread of an order of magnitude ($10^{10}$-$10^{11}$\$). However, we did
not observe any qualitative dependencies of the relevant statistical and
multifractal properties for different stocks on these quantities. Figure 1
presents the c.d.f. of the returns (a) and of the singularity spectra
$f(\alpha)$ (b) for all 100 stocks under study. While widths of the
$f(\alpha)$ spectra rather vary among the stocks, their c.d.f.'s all scale
with exponent $\simeq -3$.

\begin{figure}
\hspace{0.0cm}
\epsfxsize 14cm
\epsffile{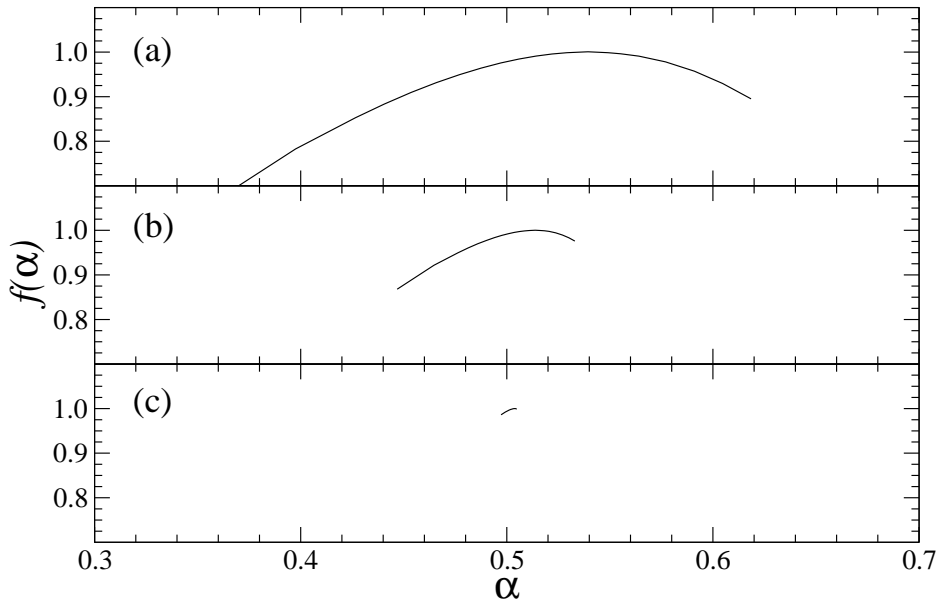}
\caption{The singularity spectra $f(\alpha)$, calculated after averaging
$h(q)$ over companies, for the original signals (a), for the reshuffled 
ones preserving distribution of the returns (b), and for the surrogates 
with both the distribution and the nonlinear correlations destroyed 
(c).}
\label{singularity}
\end{figure}

In order to separate the two possible sources of multifractality, i.e.  
the temporal correlations and the broad returns distributions, one can
destroy the temporal structure of the signals by randomly reshuffling the
corresponding time series of returns. What then remains are signals with
exactly the same fluctuation distributions but without memory. Figure 2
shows the spectra $f(\alpha)$ averaged over all the companies for the
original (a) and for the randomized (b) data. It is clear that these
spectra are significantly different from each other, with the one for the
randomized data being much narrower, i.e. much less multifractal than its
counterpart for the real data. This indicates the important role of the
temporal correlations as a source of multifractal dynamics of the stock
returns. However, also the reshuffled data is of multifractal nature,
which is in agreement with what the power-law behaviour of its
distribution (Figure 1(a)) can suggest. A contribution to multifractality
of the fat-tailed shape of the fluctuation distributions can be assessed
by comparing the $f(\alpha)$ spectrum for the randomized data with the
monofractal spectrum e.g. the one for the Fourier-phase-randomized
surrogates. The surrogates are characterized by the same linear
correlations (which do not produce multifractality~\cite{kalisky04}) as
the original data but, by construction, for long signals their
fluctuations are almost Gaussian.  By comparing Figure 2(b)  and Figure
2(c) one sees that in our case the fat-tailed distribution shape is also a
meaningful source of multifractality, though rather weaker than the
temporal correlations.

The contribution of correlations to the multifractal character of the
high-frequency stock returns one can also quantify in another way, by
considering variability of the generalized Hurst exponent
$h(q)$~\cite{kantelhardt02}. The idea is to compare the behaviour of
$h(q)$ for the real and for the randomized data, while keeping in mind 
that an entire possible difference can be attributed to the influence of
correlations. Let us define 
\begin{equation}
h_{\rm corr}(q):=h(q)-h_{\rm rand}(q), 
\end{equation}
where $h_{\rm rand}(q)$ denotes $h(q)$ for the reshuffled signals.
Since a multifractal signal is characterized by monotonously decreasing
function $h(q)$ and since richer multifractality corresponds to stronger
variability of $h(q)$, it is convenient to look at the quantity 
\begin{equation}
\Delta h:=h(q_{\rm min})-h(q_{\rm max})
\end{equation}
and its counterparts $\Delta h_{\rm rand}$ and $\Delta h_{\rm corr}$.

Contributions of the two sources to the multifractal spectra can be easily
compared by computing the ratio $R=\Delta h_{\rm corr} / \Delta h_{\rm
rand}$. The main panel of Figure 3 shows values of this ratio for all the
companies under study. What can be clearly seen is that only a few stocks
are characterized by $R \le 1$ while most of the stocks fall within the
range $1 < R < 5$. This result indicates that for the majority of stocks
the temporal correlations are responsible for the most of the width of the
multifractal spectra, leaving only the smaller remaining part for the
distributions of returns. The plot also shows that the larger values of R
are acompanied by the larger total width $\Delta h$. In fact, this
behaviour is confirmed by the inset of Fig.~3, where $\Delta h_{\rm
corr}(\Delta h)$ (full symbols) is firmly linear with slope $\sim 0.85$.
$\Delta h_{\rm rand}(\Delta h)$ (open symbols) exhibits only a small
positive slope $\sim 0.15$, suggesting that the multifractal component
related to the distributions only weakly depends on $\Delta h$.

\begin{figure}
\hspace{0.5cm}
\epsfxsize 14cm
\epsffile{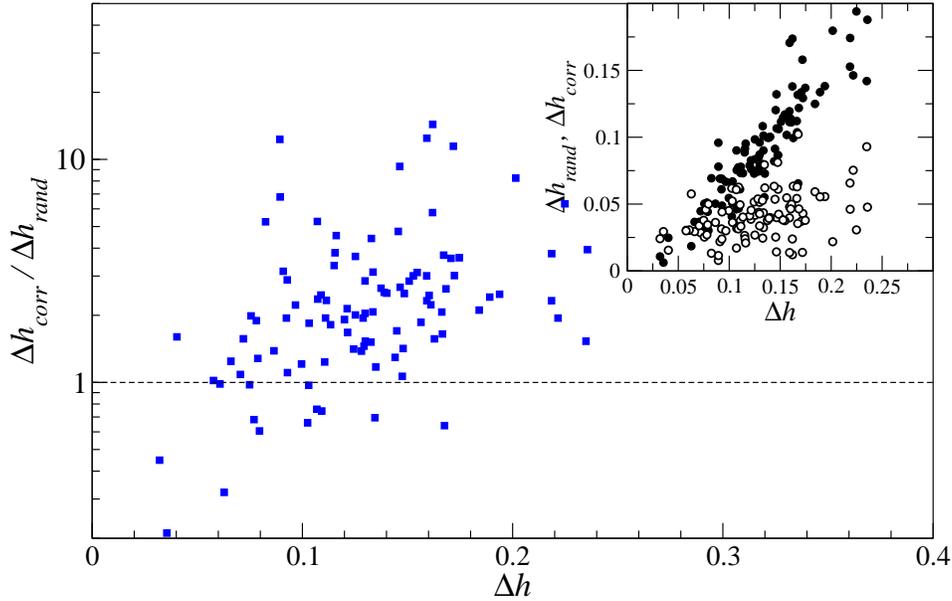}
\caption{Main: multifractal spectra's widths ratio $\Delta h_{\rm corr} / 
\Delta h_{\rm rand}$ versus $\Delta h$ for the individual companies. Inset: 
$\Delta h_{\rm corr}$ (full symbols) and $\Delta h_{\rm rand}$ (open 
symbols) vs. $\Delta h$. Weak dependence of the latter and strong linear 
relation of the former is visible.}
\label{deltah}
\end{figure}

Interestingly, the above observations apparently conflict with the results
of ref.~\cite{matia03}, where the dominating factor were the broad
probability distributions of price fluctuations. However, by looking more
carefully at the outcomes of both studies, it can be shown that they do
not necessarily contradict each other. In fact, authors of
ref.~\cite{matia03} analyzed daily returns, while we deal with the high
frequency ones for which a nonlinear temporal structure is more
complicated and more significant, with a more persistent memory (if time
lag is measured in data points), with daily patterns and with intra-daily
dependencies. It is not surprising also that in~\cite{matia03} it was the
commodities which revealed a more meaningful correlation component than
the stocks: due to the distinct nature of the commodities and the stocks,
characteristic time scales for the former are much longer than for the
latter and, thus, qualitatively, the price of the commodities sampled
daily may well correspond to a more frequently sampled data for the
stocks. Consistently, our results resemble the results for the commodities
rather than for the stocks in~\cite{matia03}.

One of the above-mentioned correlation types, the daily pattern of
volatility, which is a well-known property of all trading markets, in
itself is a subject worth bringing up in a context of the multifractal
analysis. This pattern is sometimes considered as an uninteresting
property of dynamics, thus often simply removed. Usually it is erased by
dividing each return by standard deviation or average modulus of returns
recorded at exactly the same moment of each trading day. Let us define a
daily trend for a stock $s$ by square root of the expression
\begin{equation}
\sigma_s^2(k)=(1/n_d) \sum_{j=1}^{n_d} (z_s^{(j)}(k)-<z_s^{(j)}(k)>_j)^2,
\end{equation}
where $n_d$ denotes the number of returns in a trading day and the $k$th
return of the $j$th trading day is denoted by
$z_s^{(j)}(k)=g_s((j-1)*n_d+k)$. Then a detrended return $Z_s^{(j)}(k)$
reads
\begin{equation}
Z_s^{(j)}(k)=z_s^{(j)}(k)/\sigma_s(k).
\end{equation}
This procedure e.g. leads to clearing the daily oscillations of the
volatility autocorrelation. However, it should be noted that data without
the intra-day volatility pattern present rather different statistical and
dynamical properties than the original data. To illustrate this, we
perform the multifractal analysis on the detrended (by standard deviation)  
data and compare results with the ones obtained for the non-detrended
signals. First, the detrending alters probability distribution of the
returns so that its tails are then thinner (Figure~4(a)) (see
also~\cite{wang01}); in result, it is to affect the multifractal
properties of such signals. Figure 4(b) shows the $f(\alpha)$ spectra for
the original (solid) and the detrended (dashed) data; indeed, their
distinct widths suggest that the multifractality of the latter is poorer
than the one of the former. That this effect is in a large part due to the
change of the probability distribution of the returns can be understood
from Figure 4(c), in which the $f(\alpha)$ spectra for the reshuffled
detrended signals (dashed) and the reshuffled original ones (solid) are
presented. Here the difference between the spectra is associated with the
corresponding difference between the distributions in Fig.~4(a). The
spectrum for the detrended reshuffled signals, although rather narrow, is
still of multifractal nature (it is significantly wider than its
counterpart for signals derived from purely monofractal Gaussian noise,
which resembles the spectrum in Fig.~2(c)). And what happens to that
component of multifractality, which originates from correlations? For each
individual stock, we calculated $\Delta h_{\rm corr}^{(d)}$ after the
detrending and compared it to its non-detrended counterpart $\Delta h_{\rm
corr}^{(n)}$ already displayed in Fig.~3. Outcome of this calculation can
be viewed in Figure 5. Despite the fact that $\Delta h_{\rm
corr}^{(n)}/\Delta h_{\rm corr}^{(d)}$ varies substantially from about 0.5
(detrending amplifies the component under discussion) up to 3 (detrending
suppresses it), for a predominant group of stocks there is no visible
change in $\Delta h_{\rm corr}$ (ratio $\simeq 1$). In the latter case the
detrending does not modify those correlations which are responsible for
the multifractal behaviour of signals, suggesting that for many stocks
their so-removed intra-day volatility trend only marginally participates
in multifractality.

\begin{figure}
\hspace{0.5cm}
\epsfxsize 14cm
\epsffile{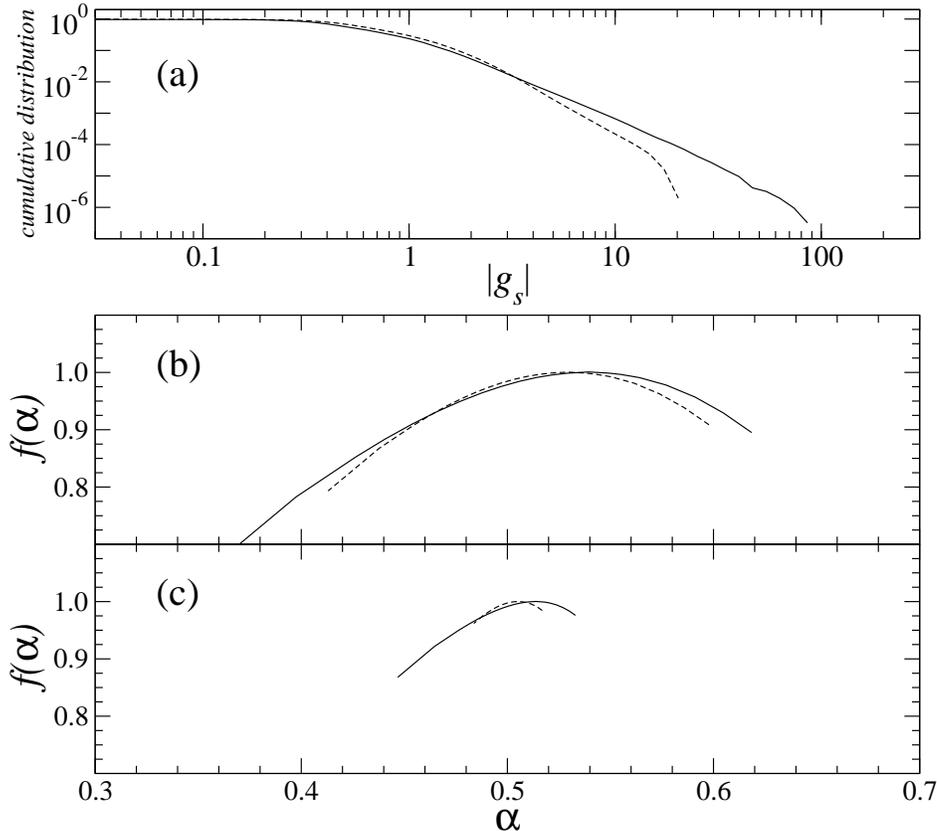}
\caption{Comparison of statistical and multifractal characteristics of 
original (solid) and detrended (dashed) signals: cumulative probability 
distributions of returns (a), singularity spectra for actual data (b), and 
singularity spectra for reshuffled data (c). Each plot presents results 
averaged over 100 stocks.}
\label{detrended}
\end{figure}

\begin{figure}
\hspace{0.5cm}
\epsfxsize 14cm
\epsffile{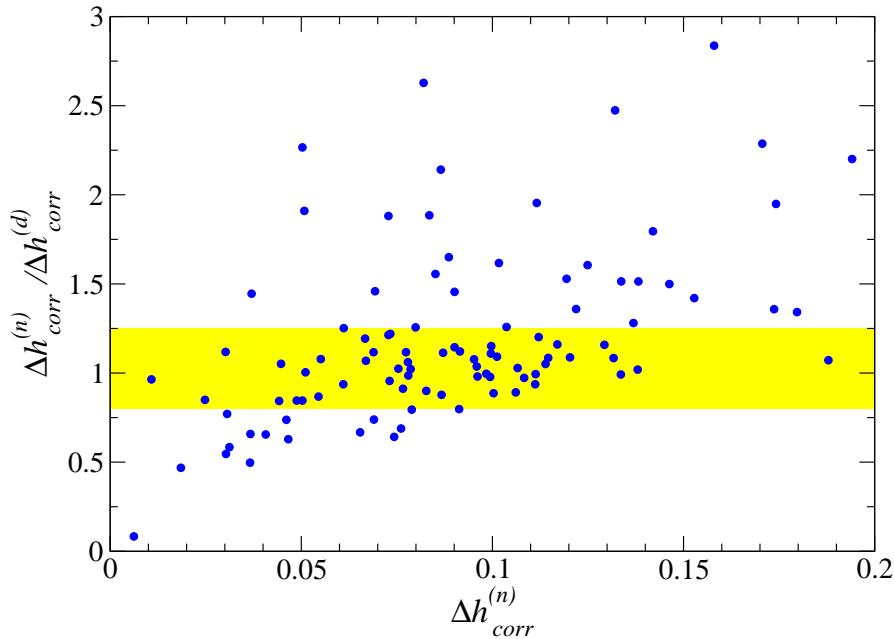}
\caption{$\Delta h_{\rm corr}^{(n)}/\Delta h_{\rm corr}^{(d)}$ 
as a function of $\Delta h_{\rm corr}^{(n)}$; symbols represent individual 
companies; shaded region (0.80,1.25) comprises over 50\% of stocks.}
\label{ratio}
\end{figure}

To summarize, we studied multifractal characteristics of the time series
of 5-minute stock returns for the 100 largest American companies. Returns
for all the stocks exhibit multiscaling, but their $f(\alpha)$ spectra
have different widths. There are two fundamental sources of
multifractality:  the fat-tailed probability distributions of returns and
the temporal correlations present in the data. The c.d.f. of returns are
similar for all the stocks and, thus, their corresponding contribution to
multiscaling is almost the same for each stock. In contrast, the temporal
correlations are a factor whose influence is company-dependent and
typically its strength is at least as large as the one for the return
distributions. For the vast majority of stocks, the temporal correlations
constitute a dominant factor spanning the spectrum. It is worthwhile to
note that similar conclusions can be drawn from results presented in our
previous paper~\cite{oswiecimka04} based on analysis of the German stocks.
This convinces us that the above-described properties cannot be thought of
as a unique attribute of the American stock market. Our results need to be
confronted with earlier study from ref.~\cite{matia03} where, for daily
stock returns, the multifractality was primarily related to the power-law
probability distributions. The daily data, however, lacks to show so much
temporal structure as the high-frequency returns do, and, in consequence,
its multifractality is likely to reflect only this one source.

\end{document}